\documentclass{article}

\usepackage{arxiv}

\usepackage[utf8]{inputenc} 
\usepackage[T1]{fontenc}    
\usepackage{hyperref}       
\usepackage{url}            
\usepackage{booktabs}       
\usepackage{amsfonts}       
\usepackage{nicefrac}       
\usepackage{microtype}      
\usepackage{natbib}
\usepackage{lipsum}
\usepackage{graphicx}
\usepackage{amsmath}
\graphicspath{ {./images/} }

\title{Bridging the Trust Gap in Crowdfunding: A Novel Expert-Based Evaluation Mechanism}

\author{
 Issam Hosni \\
  Faculty of Mathematics and Computer Science\\
  University of Tiaret\\
  Tiaret, Algeria, 14000 \\
  \texttt{issam.hosni@univ-tiaret.dz} \\
   \And
 Omar Talbi \\
  Faculty of Mathematics and Computer Science\\
  University of Tiaret\\
  Mathematics and Informatics Research Lab. (LIM)\\
  University of Tiaret\\
  Tiaret, Algeria, 14000\\
  \texttt{omar.talbi@univ-tiaret.dz} \\
}

\begin{document}
\maketitle
\begin{abstract}
Crowdfunding has emerged as a vital alternative funding source, transforming how creative projects and startups secure financing by directly connecting creators to backers. However, persistent trust issues and information asymmetry between creators and backers significantly hinder its growth and development. Existing trust-enhancement mechanisms, such as third-party endorsements and basic expert validation often lack objectivity and robustness, leaving backers vulnerable to biased signals and project failures. This paper addresses these limitations by introducing a novel trust-enhancement mechanism, referred to as \textbf{Double-Score Voting}. This approach refines expert validation systems by integrating two critical dimensions: firstly, a granular score-based vote from experts on a project's potential, moving beyond simple binary approval; and secondly, a weighted score representing the expert's credibility and level of expertise. This dual-layered evaluation provides a more nuanced, objective, and reliable assessment of project viability. The mechanism is formalised mathematically, and its practical implementation is demonstrated through \textbf{CertiFund}, a prototype crowdfunding platform developed to test and validate the concept. The findings of this study demonstrate that the Double-Score Voting mechanism can significantly mitigate information asymmetry, thereby increasing the credibility of projects and fostering a more trustworthy ecosystem for both creators and backers. 
\end{abstract}

\keywords{Crowdfunding \and Trust \and Information Asymmetry \and Signaling Theory \and Expert Validation \and Approval Voting \and Double-score Voting}

\section{Introduction}
In the aftermath of the 2008 financial crisis, crowdfunding has undergone a transformative shift, significantly impacting the financing landscape for emerging enterprises, creative projects, and social initiatives. Platforms such as Kickstarter and Indiegogo have effectively democratized access to capital by establishing direct connections between creators and a dispersed audience of potential backers [\citenum{Agrawal2011}, \citenum{Chawdhry2002}]. The global crowdfunding market is projected to experience substantial growth, underscoring its critical role in the contemporary economy [\citenum{Chawdhry2002}].\\
Despite this success, the ecosystem is plagued by a fundamental challenge: the inherent information asymmetry between project creators, who possess comprehensive knowledge, and potential backers, who have limited data with which to assess project quality and creator competence [\citenum{Belleflamme2015}, \citenum{Brandl2022}]. This information gap fosters distrust, which is a critical barrier in online transactions where face-to-face interactions are absent [\citenum{Chen2022}, \citenum{Moysidou2020}]. Consequently, high-quality projects may remain unfunded, while backers risk supporting projects that are susceptible to failure or fraud.\\
To mitigate these issues, platforms have adopted mechanisms derived from Signaling Theory [\citenum{Connelly2011}]. Creators signal quality through detailed descriptions, media content, and regular updates [\citenum{Li2024}]. Third-party endorsements, such as backer comments and social media shares, also serve as signals, though they are often susceptible to social proof and herding biases [\citenum{Li2024}, \citenum{Roy2021}].\\
A more structured approach has been the introduction of expert validation systems. However, current implementations are often rudimentary, relying on a binary "yes/no" or "approval/rejection" voting method [\citenum{Hu2024}]. This approach is restrictive, fails to capture the nuances of an expert's opinion, and treats all expert opinions as equal, regardless of their specific expertise or credibility. This leads to a critical gap: the lack of an objective, robust, and transparent mechanism for expert evaluation.
To address this gap, this paper seeks to answer the following research questions:
\begin{itemize}
    \item \textbf{RQ1:}	How can expert validation systems in crowdfunding platforms be designed to ensure greater objectivity and robustness in project assessment?
    \item \textbf{RQ2:} What are the practical challenges that face the integration of a robust and transparent voting mechanism within an existing crowdfunding platform framework?
    \item \textbf{RQ3:} What is the impact of such a structured, more rigorous expert validation system on mitigating information asymmetry and fostering trust between creators and backers?
\end{itemize}
To answer these questions, we adopt a constructive research methodology. First, drawing from voting theory and critiques of existing systems, we design and formalize a novel mechanism: the Double-Score Voting system \textbf{(addressing RQ1)}. Second, to establish its practical feasibility, we detail its implementation within CertiFund, a proof-of-concept crowdfunding platform, outlining architectural and workflow solutions to key integration challenges \textbf{(addressing RQ2)}. Finally, we analyze the theoretical impact of this mechanism on the crowdfunding ecosystem by linking its design features directly to the core problems of information asymmetry and trust \textbf{(addressing RQ3)}. The remainder of this paper is structured to follow this methodological approach.

\section{Background and Related Work}
\subsection{Trust and Signaling in Crowdfunding}
Trust is paramount for the success of any online exchange [\citenum{Lee2005}]. In the context of crowdfunding, this critical factor (trust) emerges in two forms: calculus-based trust, derived from a rational evaluation of signals such as project descriptions or endorsements, and relationship-based trust, established through interactions between creators and backers [\citenum{Li2024}]. The primary challenge confronting platforms is to furnish credible signals that can surmount initial distrust and mitigate information asymmetry.\\
Signaling theory offers a framework for understanding how creators convey unobservable qualities. Research has demonstrated a positive correlation between effective signals, such as creator experience, project preparedness, and third-party endorsements, and funding success [\citenum{Li2024}, \citenum{Vismara2017}]. Nevertheless, the efficacy of these signals is frequently constrained. Visual cues can be manipulated, and endorsements from non-experts can create misleading social proof, leading to a "lemons problem" where low-quality projects crowd out high-quality ones [\citenum{Pan2008}].
\subsection{The Limits of Existing Expert Validation}
To address the issue of unreliability associated with conventional signals, certain platforms have adopted an approach that involves the integration of expert validation. The evaluation process entails a panel of experts who assess projects based on their feasibility and the competence of the creator. This approach has been demonstrated to provide a more robust and objective signal in comparison to the endorsements provided by the general public [\citenum{Hu2024}, \citenum{Petit2021}].
Nevertheless, the prevailing method of binary (yes/no) voting is critically flawed. This approach simplifies a complex judgment into a binary choice, resulting in a substantial loss of information. Additionally, this method disregards the recognized distinction between experts and novices, thereby overlooking the variance in the depth and reliability of their respective knowledge. This fundamental limitation prevents existing expert validation systems from achieving their full potential in fostering backer confidence and reliably signaling project quality.

\section{The Double-Score Voting Mechanism: A Formal Design}
In response to the need for a more objective and robust evaluation system, we propose the Double-Score Voting mechanism. This design moves beyond simplistic binary choices to incorporate the principles of nuanced evaluation and credibility weighting, directly addressing the flaws identified in the existing literature.
\subsection{From Binary to Score-Based Voting: Capturing Nuance}
A simple approval voting system in our context would allow an expert to select one or more recommendation levels to assess a particular project (e.g., "Highly Recommended", "Recommended", ...). This is logically inconsistent... A superior approach is \textbf{score voting}, where each expert distributes a total of 100 points (or 100\%) across the possible recommendation levels... This captures the expert's precise degree of confidence and uncertainty, providing a far more \textbf{robust} evaluation than a simple binary choice.
\subsection{Incorporating Expert Credibility: Ensuring Objectivity}
The key innovation of our mechanism is the introduction of a second score: the expert's credibility, or expertise weight (\( \lambda_e \)). This second layer ensures that the opinions of more qualified and credible experts have a greater impact on the outcome. This design choice directly enhances the system's objectivity by moving away from a simple majority opinion towards a weighted, meritocratic judgment.
\subsection{Formalization of the Mechanism}
To fully specify our proposed design, we formalize the Double-Score Voting algorithm as follows:\\
For each recommendation level \( l \in \{HNR, NR, R, HR\} \), the score \( S_l \) in a double-weighted recommendation voting system is computed as:

\[
S_l = \sum_{e \in E} \lambda_e \cdot w_e^l
\]
Where:
\begin{itemize}
    \item \( E \): Set of experts evaluating the project.
    \item \( \lambda_e \): Expertise weight of expert \( e \), where \( \lambda_e \geq 0 \).
    \item \( w_e^{HNR}, w_e^{NR}, w_e^R, w_e^{HR} \): Weights assigned by expert \( e \) to ``highly not recommended'' (HNR), ``not recommended'' (NR), ``recommended'' (R), and ``highly recommended'' (HR), with \( w_e^{HNR} + w_e^{NR} + w_e^R + w_e^{HR} = 1 \).
\end{itemize}

The final recommendation is determined by the level with the highest aggregated score:

\[
\text{Final Recommendation} = \arg\max_{l \in \{HNR, NR, R, HR\}} S_l
\]

\section{Implementation and Feasibility Analysis: The CertiFund Prototype}
To validate the practical viability of the proposed mechanism (RQ2), we developed \textbf{CertiFund}, a proof-of-concept platform that serves as a research "artefact". This section details how the theoretical design was translated into a functional system, outlining the architectural solutions to key integration challenges and illustrating the complete expert validation workflow. 
\subsection{System Architecture and Technology Choices}
The foundation of CertiFund is a modern technology stack, chosen to ensure performance, maintainability, and a robust implementation environment. The system follows a layered, monolithic architecture with a decoupled frontend and backend communicating via a RESTful API. 
\begin{itemize}
    \item \textbf{Backend}: Developed in Go with the Echo framework for its high performance and concurrency handling.
    \item \textbf{Frontend:} Built with Next.js (React) and TypeScript, providing a responsive and type-safe user interface.
    \item \textbf{Database:} PostgreSQL was selected for its reliability and, crucially, its support for advanced custom functions, which proved instrumental in solving the logic placement challenge.
    \item \textbf{Automation:} The system relies on containerization with Docker and task scheduling via standard Linux cron jobs.
\end{itemize}
This architecture provided the necessary tools to address the primary practical challenges of implementation. To access the complete codebase of the platform, follow this link: \hyperlink{https://github.com/issam-hub/CertiFund}{https://github.com/issam-hub/CertiFund}.
\subsection{From Theory to Practice: The Expert Validation Workflow}
\begin{figure}[h]
\centering
\includegraphics[width=0.75\linewidth]{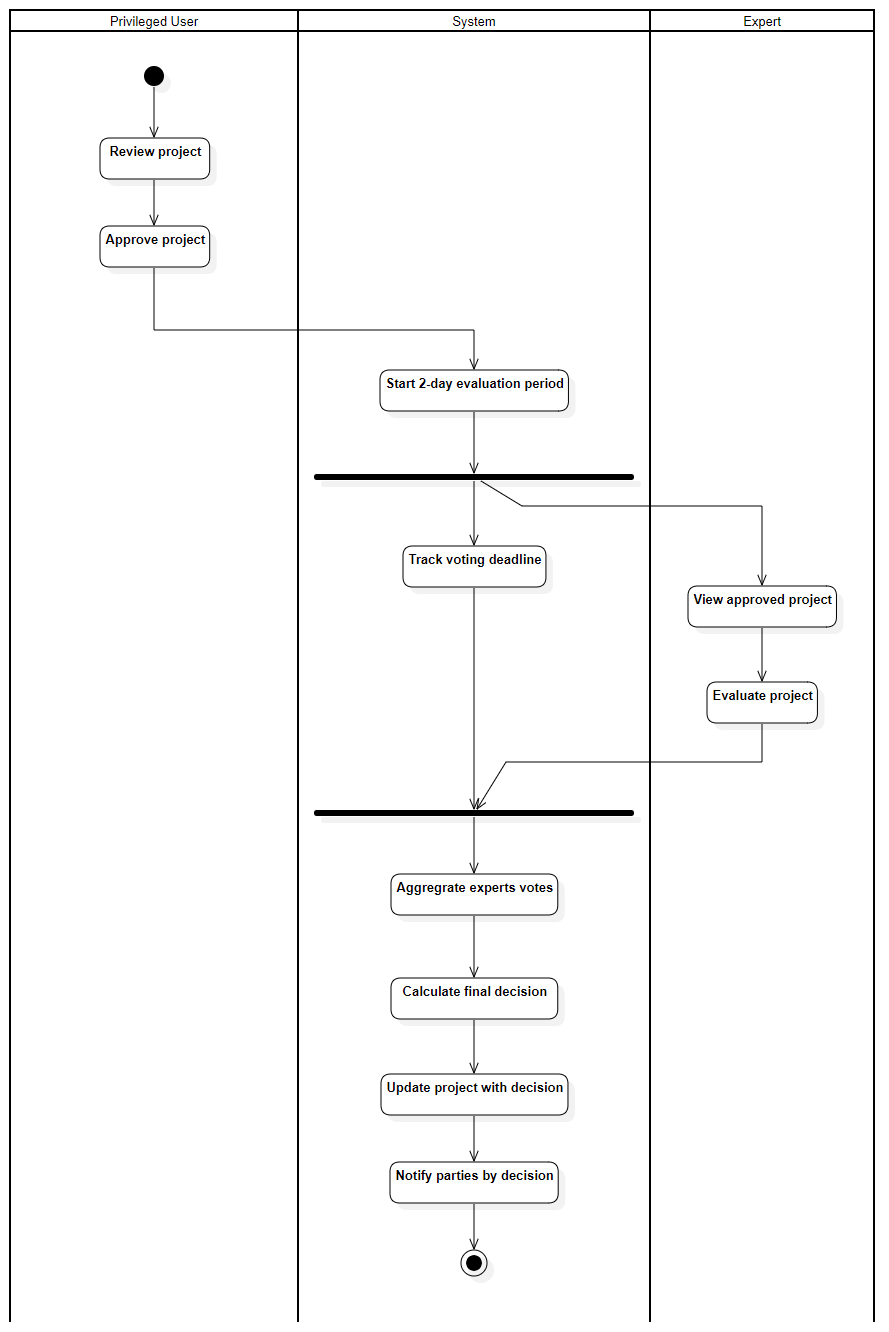}
\caption{Activity diagram presenting the Expert Validation workflow on CertiFund, from project approval to the final decision notification.}
\label{fig:1}
\end{figure}

The core of the implementation lies in the end-to-end workflow for expert validation. This process demonstrates how architectural choices and user interface design work in concert to bring the Double-Score Voting mechanism to life. The workflow is best understood through the sequence of user interactions and automated processes, as illustrated in the activity diagram in Figure \ref{fig:1}
\\ The workflow, as depicted, proceeds in four key stages:
\begin{itemize}
    \item \textbf{Stage 1: Triggering the Evaluation:}
The process begins when a privileged user (e.g., a platform administrator) approves a creator's project. This action does not make the project public but instead initiates a fixed 2-day expert evaluation period and notifies all relevant experts based on their declared specializations.
    \item \textbf{Stage 2: Nuanced Vote Collection:}
During the evaluation period, notified experts review the project details. They then submit their assessment using a dedicated interface, as shown in Figure \ref{fig:2}. This form is the practical embodiment of our mechanism's "score voting" principle. Instead of a simple checkbox, it features sliders or input fields, allowing experts to distribute 100 percentage points across the recommendation levels. This design solves the challenge of collecting nuanced input by providing an intuitive yet powerful tool for expressing a detailed opinion.
    \begin{figure}[h]
    \centering
    \includegraphics[width=0.6\linewidth]{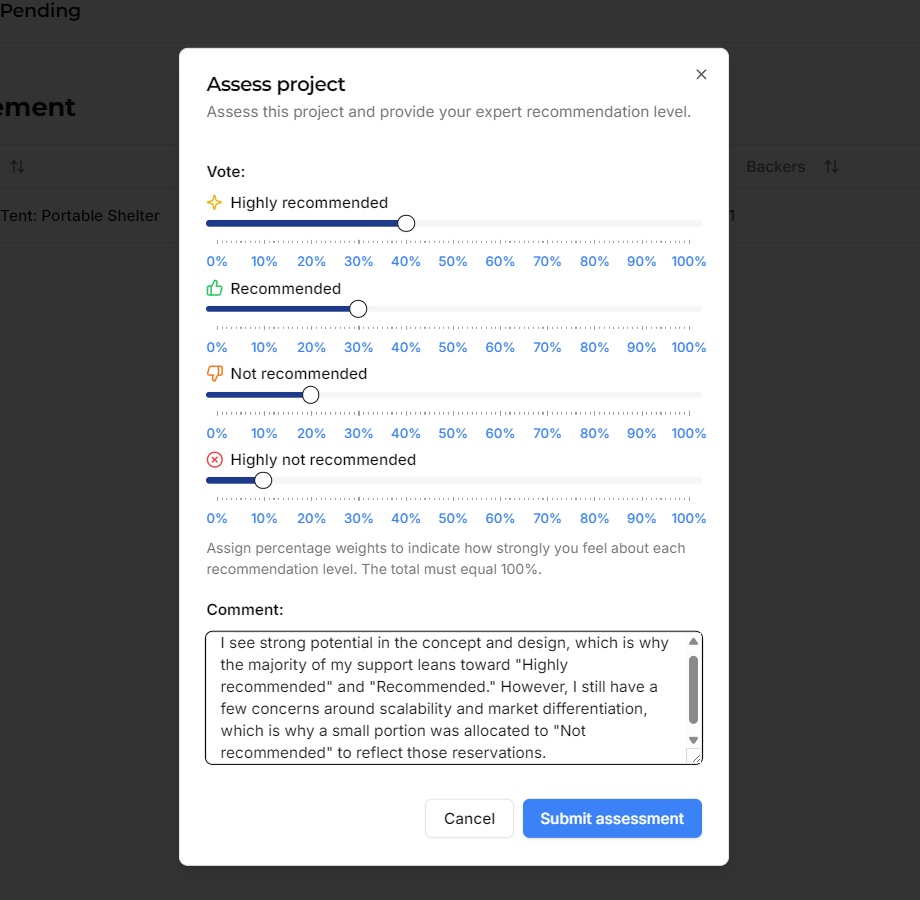}
    \caption{The Double-Score Voting interface on CertiFund. This form allows experts to provide a granular, score-based assessment of a project.}
    \label{fig:2}
    \end{figure}
    \item \textbf{Stage 3: Automated Vote Aggregation:}
A critical challenge is ensuring the aggregation is timely and reliable. This is handled by a Linux cron job that runs periodically. At the end of the 2-day window, the cron job triggers the calculate\_expert\_decision function within the PostgreSQL database. This function, which contains the core Double-Score Voting algorithm, aggregates all submitted votes for the project, weighting each according to the expert's credibility score (\(\lambda_e\))). Placing this logic directly in the database ensures performance, atomicity, and decouples the complex calculation from the main backend application.
    \item \textbf{Stage 4: Displaying the Outcome:}
Once the calculation is complete, the final aggregated recommendation (e.g., "Recommended by Experts") is stored and displayed prominently on the project's public page. This outcome serves as a \textbf{high-quality}, \textbf{institutionalized signal} for potential backers, completing the workflow.
\end{itemize}
The successful implementation of this end-to-end process within CertiFund demonstrates that the practical challenges of architecture, automation, and user experience are surmountable, thus validating the feasibility of the Double-Score Voting mechanism.

\section{Analysis of Impact on the Crowdfunding Ecosystem}
Having established the design and feasibility of the Double-Score Voting mechanism, we now analyze its potential impact on the crowdfunding ecosystem. The mechanism's features are designed to fundamentally improve the quality of information available to backers, thereby strengthening the foundations of trust.
\subsection{Mitigating Information Asymmetry through High-Quality Signaling}
The primary impact of the system is its ability to generate a high-quality, institutionalized signal. Unlike subjective backer comments, the final recommendation (e.g., "Recommended by Experts"), often accompanied by the experts' anonymous comments, is a credible piece of information that directly reduces the knowledge gap between creators and backers. Backers are no longer left to interpret weak or potentially biased signals; they receive a clear, aggregated, and objective assessment of project quality. Given that, this signal-based method directly addresses the "lemons problem" [\citenum{Pan2008}] and reduces the impact of information asymmetry [\citenum{Hu2024}, \citenum{Li2024}].
\subsection{Fostering Institutional and Calculus-Based Trust}
Trust is fostered in two ways. First, the \textbf{transparency} of the mechanism itself builds confidence. Knowing that projects are vetted through a rigorous and objective process enhances trust in the platform as a whole (institutional trust). Second, a positive/negative expert recommendation increases \textbf{calculus-based trust/distrust} in the specific project. Backers can rationally decide to place their trust (and funds) in a project because a panel of credible experts has collectively endorsed its viability, or they can save themselves from a potential fraud [\citenum{Moysidou2020}]. This structured process provides a defensible reason to trust, moving beyond gut feelings or social herding [\citenum{Li2024}].
\section{Conclusion and Future Work}
This paper identified a critical weakness in current crowdfunding trust mechanisms—the lack of robust and objective expert validation—and proposed a novel solution: the \textbf{Double-Score Voting} mechanism. By formalizing this algorithm and demonstrating its implementation in the CertiFund prototype, we have laid the groundwork for a new generation of trust-driven crowdfunding platforms.\\
Future research should proceed in several directions. First, an \textbf{empirical study} on a live platform is needed to validate the real-world impact of our mechanism. Second, research into \textbf{automated} and \textbf{dynamic} calculation for the expert score, potentially using machine learning models (leveraging LLM models), could further enhance the objectivity of the \(\lambda_e\) parameter. Finally, exploring the integration of this mechanism with \textbf{decentralized} technologies like \textbf{blockchain} could pave the way for fully transparent and immutable trust systems in the future.

\bibliographystyle{abbrv}
\bibliography{bib}

\begin{thebibliography}{10}

\bibitem{Agrawal2011}
A.~K. Agrawal, C.~Catalini, and A.~Goldfarb.
\newblock The geography of crowdfunding.
\newblock Working Paper 16820, National Bureau of Economic Research, February 2011.

\bibitem{Belleflamme2015}
P.~Belleflamme, N.~Omrani, and M.~Peitz.
\newblock The economics of crowdfunding platforms.
\newblock {\em ERN: Information Asymmetry Models (Topic)}, 2015.

\bibitem{Brandl2022}
F.~Brandl and D.~Peters.
\newblock Approval voting under dichotomous preferences: A catalogue of characterizations.
\newblock {\em Journal of Economic Theory}, 205:105532, 2022.

\bibitem{Chen2022}
D.~Chen, C.~Huang, D.~Liu, and F.~Lai.
\newblock The role of expertise in herding behaviors: evidence from a crowdfunding market.
\newblock {\em Electron. Commer. Res.}, 24:155--203, 2022.

\bibitem{Connelly2011}
B.~L. Connelly, S.~T. Certo, R.~D. Ireland, and C.~R. Reutzel.
\newblock Signaling theory: A review and assessment.
\newblock {\em Journal of Management}, 37(1):39--67, 2011.

\bibitem{Hu2024}
J.~Hu, W.~You, and B.~Chen.
\newblock The impact of experts' voting on the fundraising performance of crowdfunding projects.
\newblock In {\em Wuhan International Conference on E-Business}, 2024.

\bibitem{Chawdhry2002}
C.~P. k., M.~Marcelo, and W.~M. andre.
\newblock Strategies for trust and confidence in b2c e-commerce.
\newblock 2002.

\bibitem{Lee2005}
B.-C. Lee, L.~Ang, and C.~Dubelaar.
\newblock Lemons on the web: A signalling approach to the problem of trust in internet commerce.
\newblock {\em Journal of Economic Psychology}, 26:607--623, 2005.

\bibitem{Li2024}
Q.~Li and N.~Wang.
\newblock Fundraiser engagement, third-party endorsement and crowdfunding performance: A configurational theory approach.
\newblock {\em PLOS ONE}, 19, 2024.

\bibitem{Pan2008}
Y.~ming Pan.
\newblock Improving the cyber 'lemons' market with a trust-intermediary in e-commerce: Model and case study.
\newblock {\em 2008 4th International Conference on Wireless Communications, Networking and Mobile Computing}, pages 1--6, 2008.

\bibitem{Moysidou2020}
K.~Moysidou and J.~P. Hausberg.
\newblock In crowdfunding we trust: A trust-building model in lending crowdfunding.
\newblock {\em Journal of Small Business Management}, 58:511 -- 543, 2020.

\bibitem{Petit2021}
A.~Petit and P.~Wirtz.
\newblock Experts in the crowd and their influence on herding in reward-based crowdfunding of cultural projects.
\newblock {\em Small Business Economics}, 58:419 -- 449, 2021.

\bibitem{Roy2021}
S.~Roy.
\newblock Theory of social proof and legal compliance: A socio-cognitive explanation for regulatory (non) compliance.
\newblock {\em German Law Journal}, 22:238 -- 255, 2021.

\bibitem{Vismara2017}
S.~Vismara.
\newblock Signaling to overcome inefficiencies in crowdfunding markets.
\newblock {\em IRPN: Innovation \& Finance (Topic)}, 2017.

\end{thebibliography}

\end{document}